\documentstyle[draft]{l-aa}

\thesaurus{08.01.1 08.05.3 08.09.3 08.18.1 02.04.2}
\newcommand{\etal}{{\it et al.}}
\newcommand{\kms}{\mbox{\ km\ s$^{-1}$}}
\newcommand{\msunyr}{\mbox{M$_{\odot}$\thinspace yr$^{-1}\;$}}
\newcommand{\msun}{\mbox{$M_{\odot}\;$}}
\newcommand{\mstar}{\mbox{$M_{\ast}\;$}}
\newcommand{\rsun}{\mbox{$R_{\odot}\;$}}
\newcommand{\rstar}{\mbox{$R_{\ast}\;$}}

\newcommand{\mnras}{{\sl MNRAS}}
\newcommand{\apj}{{\sl ApJ}}

\newcommand{\apjs}{{\sl ApJS}}

\newcommand{\aanda}{{\sl A\&A}}

\newcommand{\mdot}{\mbox{$\stackrel{.}{\textstyle M}$}}
\newcommand{\ltappeq}{\mathrel{\hbox{\rlap{\hbox{\lower4pt\hbox{$\sim$}}}\hbox{$<$}}}}
\newcommand{\gtappeq}{\raisebox{-0.6ex}{$\,\stackrel{\raisebox{-.2ex}{$\textstyle >$}}{\sim}\,$}}

\title{On the possibility that rotation causes latitudinal abundance
variations in stars} 
\author{John M. Porter}
\institute{Astrophysics Research Institute, 
School of Engineering, Liverpool John Moores University,
Byrom Street, Liverpool L3 3AF, UK
\thanks{Current address : Astrophysics Research Institute, 
Liverpool John Moores University, Twelve Quays House, Egerton Wharf,
Birkenhead. L41 1LD } \\
(email : {\tt jmp@astro.livjm.ac.uk}) 
}

\date{Recieved October 6; accepted October 22, 1998}

\begin{document}
\maketitle

\begin{abstract}

The effect of rotation of a star on the distribution of chemical
species in radiative zones is discussed. Gravity darkening generates a
large radiative force on heavy element ions which is
directed toward the equatorial plane. 
Taking iron as an example, 
it is shown that this force may produce drift velocities similar to,
and larger than, the typical velocities of bulk motion due
to meridional circulation. 
This potentially allows large chemical abundance inhomogeneities to
build up across a meridian
over the lifetime of the star -- particularly near the equatorial plane. 
This enhancement may be significantly reduced if 
the mass loss of the star is strongly metallicity dependent, in which
case the mass-loss rate may be enhanced in the equatorial plane.


\end{abstract}

\section{Introduction}

Abundance anomalies are observed in many different types of star --
the photospheres of stars with effective temperatures
hotter than $\sim 6000$K may have undergone chemical separation 
(Michaud 1970).
It is generally accepted that 
this may be explained via the force on atoms
by radiation: if this is larger than gravity, then the element will
diffuse toward the surface, else it will settle.
The force due to radiation pressure on elements in a star has been
discussed at length by e.g.
Michaud \etal\ (1976)
Vauclair \etal\ (1978), Aleican \& Artru (1990) and Gonzalez \etal\ (1995). 
Abundance anomalies have been discussed in a variety of chemically
peculiar stars 
eg. Am and Fm stars (Charbonneau \& Michaud 1991, Alecian 1996), 
$\lambda$~Bootis stars (Michaud \& Charland 1986) and magnetic Ap-Bp
stars (Alecian \& Vauclair, 1981, Michaud \etal\ 1981). 
When the star is also rotating, the surface abundance anomalies
generated by radiation pressure may be destroyed by the
action of turbulent motions within the star (Charbonneau 1992, Talon
\& Zahn 1997).
These arise from shear motions
created by the meridional motions which arise due to gravity
darkening (see eg. von Zeipel 1924, Tassoul \& Tassoul 1982, Zahn 1992)

There is some observational evidence for 
{\em latitudinal} abundance variations of heavy elements 
in planetary nebulae (Balick \etal\ 1994).
This is possibly explained by more highly processed material being
ejected faster but at later times then the rest of the shell.
However, the observations raise an interesting question: it is
possible that there are stars which are chemically inhomogeneous in
both radial and latitudinal directions?

Here a non-magnetic mechanism which generates latitudinal abundance
variations is suggested  
and investigated. It is found that a significant latitudinal radiative
acceleration on heavy elements exists in rotating stars due to gravity
darkening (von Zeipel 1924), which causes ionic species to diffuse
toward the equatorial plane.
Competing with this equatorward drift are the meridional circulation
currents, and turbulence generated via shear motions. 
The aim of this paper is solely to assess whether any latitudinal
drift of heavy elements may occur in rotating stars.

The transport of chemicals is discussed in \S2.
In \S3 the radiative force due to gravity darkening is calculated and
in \S4 the drift velocity of metals around an equipotential is
derived. Numerical results of the diffusion velocity in
stellar envelopes are presented in \S5, and by considering the
relevant timescales, 
\S6 identifies the important regions within a star for diffusion to occur. 
A discussion is presented in \S7, and conclusions
given in \S8. 

\section{Transport of chemicals}
The conservation equation for a trace element of concentration $c$ is
\begin{equation}
\rho\frac{\partial c}{\partial t} +
\nabla.\left\{ -\rho {\mathbf D}.\nabla c + \rho{\mathbf u}c
+ \rho{\mathbf v}c \right\} = 0
\end{equation}
where ${\mathbf D}$ is the diffusivity tensor, ${\mathbf u}$ is the
global motion of the fluid (in this case the meridional motions
generated via the rotation of the star), and
${\mathbf v}$ describes any motion affecting only the contaminant.
The concentration $c$ is defined as the number density of ions compared to
the number density of protons.
Hereafter, the star is assumed to be axisymmetric, so that using a
spherical coordinate system, for any $X$, $dX/d\phi = 0$. 
Also, for rotating stars, the equipotential is not defined by $r =
$constant. Therefore the ``horizontal'' direction is taken to be
perpendicular to the equipotential, and the ``vertical'' direction to be
parallel to equipotential.

Following the work of Chaboyer \& Zahn (1992) and Zahn (1992),
Charbonneau (1992) examined the effect of anisotropic turbulent
diffusion on the distribution of $c(r, \theta)$ (Zahn 1975, Tassoul
\& Tassoul 1983, Zahn 1987), and found
that if the horizontal Reynold number is $R_H < 1$ then the equation
above loses its bi-dimensional behaviour, and becomes one dimensional
ie. $c(r,\theta) \rightarrow c(r)$. Essentially the horizontal
turbulence becomes vigorous enough that it may redistribute the
concentration of the trace element faster than meridional motions can
generate an anisotropy.
This also makes the angular momentum independent of $\theta$ and gives
rise to ``shellular'' rotation of the star (Zahn 1992).

In the current study, however, the velocity field ${\mathbf v}$ may be
non-zero due to the meridional forces on trace elements arising from
the anisotropy of the radiation flux (see \S2).
To assess its likely effect in the presence of bulk meridional
motions and diffusion, the ratio of the (turbulent) diffusive term to
the radiation driven advective term across an equipotential ${\mathbf s}$
in eq.1 is examined:
\begin{equation}
{\cal R} = \left. \frac{\nabla(\rho {\mathbf D}.\nabla c)}{\nabla
(\rho{\mathbf v}c)} 
\right|_s \sim
\left[\frac{D_H }{v_\theta }\right]
\frac{1}{r}
h(c)
\end{equation}
where $D_H$ is the horizontal diffusion coefficient, the 
extra velocity component is ${\mathbf v} = (0, v_s, 0)$ and
$h(c)$ is a function containing logarithmic derivatives of the
concentration with respect to $\theta$.
This is only
an approximation as the unit vectors along an equipotential ${\mathbf
\hat{s}}$ and $\hat{\mathbf \theta}$ are not 
parallel.
However, this expression will suffice for the current study. 
Also it should be noted that there is no term in the angular momentum
evolution equation equivalent to the radiation term
$\nabla(\rho {\mathbf v}c)$ in eq.1. Therefore, shellular rotation 
will
still be achieved in the star, although the distribution of the trace
element may be anisotropic.

The discussion of turbulent diffusion coefficients $D_H$ has
a long history, and the simple approximation due to Chaboyer \&
Zahn (1992) is adopted here:
\begin{equation}
D_H = \frac{ |r u_r|}{C_H}
\end{equation}
where $C_H$ is a number of order unity and $u_r$ is the radial
component of the velocity of the bulk meridional circulation.
With this replacement eq.2 becomes 
\begin{equation}
{\cal R} \sim \left[\frac{u_r}{v_s C_H}\right]
h(c)
\end{equation}
The main controlling parameter here is the ratio of the radial
component $u_r$ of the bulk meridional flow and the velocity $v_s$ of the
trace element due to the radiation as it slips through the plasma.

If ${\cal R}>1$ then the turbulent diffusion term will dominate the
movement of trace elements, and so the star will become chemically
homogeneous on equipotentials. However, if ${\cal R} < 1$ then the
distribution of the elements is determined by the extra drift velocity 
$v_\theta$. 
The criterion adopted in this paper for chemically inhomogeneous stars
is that the extra drift velocity $v_s$ is larger than the typical
velocity of meridional motions $u_r$.
In a recent numerical study of the a 20\msun star, Urpin \etal\
(1996) found that for high rotational velocities, the circulation
velocity is less than $3\times 10^{-5}$cm s$^{-1}$. This value is
significantly lower than that from Eddington-Sweet theory, which fails
close to the surface of the star. 
If velocities of $v_s \gtappeq 10^{-5}$cm~s$^{-1}$ are produced due
to the effects of rotation, then the star may develop metal-rich and
metal-poor regions on the same equipotential.

\section{Radiation forces}

\subsection{Flux around the star}
Assuming that the potential of the outer parts of a rotating star may
be represented 
using the Roche approximation, the radius of equipotential surfaces
$r$ is
dependent on the angle to the pole $\theta$, and is found by the
solution of
\begin{equation}
\frac{r(\theta)}{r_{\rm pole}} - \frac{\omega^2}{2GM} {\rm sin}^2\theta
\left( \frac{r(\theta)}{r_{\rm pole}} \right)^3 = 1,
\end{equation}
where $\omega$ is the angular velocity, $r_{\rm pole}$ is the radius
in the polar direction, and $M$ is the mass contained
within the equipotential surface.
This has been solved for $r(\theta)$ geometrically by Collins \&
Harrington (1966). 
Deep within the star the Roche approximation will be incorrect,
although for the present discussion, it is adequate.
Indeed the region in which it is possible to generate
latitudinal metallic abundance anisotropies is found, {\it a posteriori},
to be the outer parts of the envelope and so
$M \approx \mstar$, and eq.5 will be a good representation.

The flux of the star $f$ is proportional to the local gravity (von Zeipel
1924):
\begin{equation}
f \propto \nabla
\left(-\frac{GM}{r(\theta)}
- \frac{\omega^2 r(\theta)^2 {\rm sin}^2\theta}{2}\right).
\end{equation}
Along
the surface of any equipotential the local flux $f$ is higher at the pole
than the equator (gravity darkening).
This difference between pole and equator is the root cause of the
latitudinal variation of the metallic abundance. 
It is noted that this gravity darkening actually causes the meridional
circulation, else energy would not be conserved in the star. However,
the change in flux is not destroyed by the motions.

\subsection{Radiation force on a trace element}
The force on ions due to radiation $g_i^{\rm rad}$ is dependant on the
gradient of the flux passing through that point, i.e. $g_i^{\rm
rad}~\propto\nabla f$. To facilitate the calculation of the force
along an equipotential 
the ratio of the forces in the radial and latitudinal directions is
now calculated. In the outer parts of the star, the luminosity is a
constant, and hence the flux varies with distance $r$ from the centre
of the star as $f\propto r^{-2}$. As the radiative force is
proportional to $\nabla f$, then the ratio of the latitudinal to radial
components is
\begin{equation}
\frac{g^{\rm rad}_{i, s}}{g^{\rm rad}_{i, r}} \approx
\left( \frac{df}{ds} \right) \left( \frac{df}{dr} \right)^{-1}= 
-\frac{r}{2} \frac{d{\rm ln}f}{ds}
\end{equation}
where $s$ is the distance along an equipotential. 
Again, note this is only
an approximation as the unit vectors ${\mathbf \hat{s}}$ and ${\mathbf
\hat{r}}$ are not 
orthogonal.
This ratio is shown in fig.1 for different rotation rates. 
As can be seen it is typically $0.1\rightarrow 1$,
and therefore, perhaps surprisingly, the radiative force around an
equipotential is typically similar to the radial radiative acceleration. 

\begin{figure}
\begin{picture}(100,270)
\put(0,0){\includegraphics{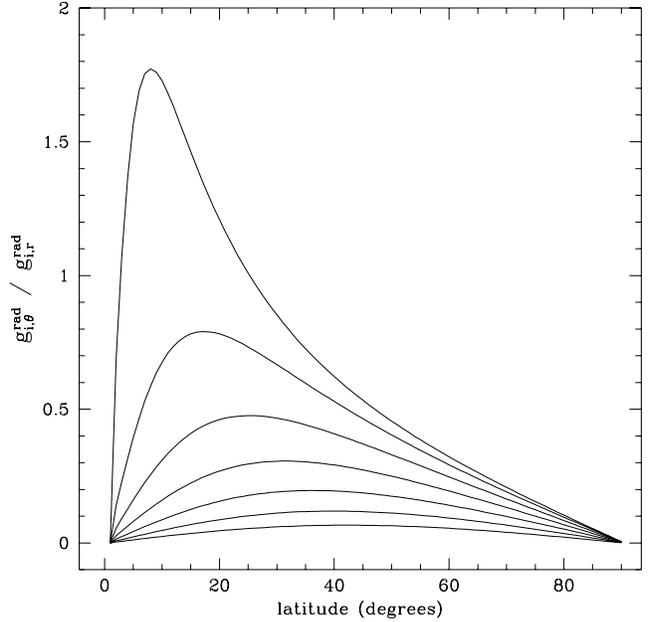}}
\end{picture}
\caption{Ratio of latitudinal to radial radiative force defined from eq.4.
The lines correspond to a rotation of 0.3, 0.4, 0.5, 0.6, 0.7, 0.8 and
0.9 of break-up velocity, where the uppermost line corresponds to the
highest rotation, decreasing downwards.}
\end{figure}

Michaud \etal\ (1976) see also Alecian \& Artru (1990) have discussed
the radiative force due to lines on heavy elements. 
Their expressions, along with eq.7, lead to the latitudinal force on 
on a trace element 
(with a negligible gradient in the concentration of the ions) of
\begin{equation}
g^{\rm rad}_{i, s}(c \rightarrow 0) = 
-5.574\times 10^7 \frac{T_{e\;4}^4}{A_i T_4} \left(
\frac{\rstar}{r} \right)^2 \Phi_i 
\left( \frac{r}{2} \frac{d{\rm ln}f}{ds} \right)
\end{equation}
where $T_4$ and $T_{e\;4}$ are the temperature and effective
temperature in units of $10^4$K respectively, $r$ is the radius and
\rstar is the stellar radius. The atomic parameters are enveloped in
$\Phi_i$, and $A_i$ is the atomic mass of the element $i$.

With the inclusion of a non-zero concentration of ions the driving
lines become saturated, and the line force is reduced from eq.8:
\begin{equation}
g^{\rm rad}_{i, s}(c \neq 0) = 
g^{\rm rad}_{i, s}(c \rightarrow 0) 
\left( \frac{1}{1 + \frac{c}{c_0} } \right)^{1/2}
\end{equation}
(Alecian 1985). The reference concentration $c_0$ is
\begin{equation}
c_0 = 9.83\times 10^{-23} \bar{\kappa} n_e T^{-1/2} \Psi_i^2
\end{equation}
where $\bar{\kappa}$ is the Rosseland mean opacity, $n_e$ is the
number density of electrons and $\Psi_i$ is an atomic parameter
containing line broadening effects.

\section{Drift velocity}
To calculate a typical order of magnitude of the drift velocity
${\mathbf v}$ between
the ions and rest of the plasma, the radiation force will be
approximately balanced by the retarding frictional force.
The diffusion velocity is
\begin{eqnarray}
{\mathbf v} & = & D_{12} \left( -\nabla {\rm ln}c +
A_i \frac{m_p {\mathbf g^{\rm rad}_i}}{kT} \ + \right. \nonumber \\
 &  & \left. \ \ \ \ \ \ \ \ \ \ \ \ \ (1-2A_i +Z_i)\frac{m_p {\mathbf g}}{kT} 
-\alpha_T \nabla {\rm ln}T \right).
\end{eqnarray}
Here, $m_p$ is the mass of a proton, $k$ is Boltzmann's constant, 
$Z_i$ is the charge of the element $i$ in units of the fundamental
charge $e$.
The thermal diffusion coefficient is $\alpha_T$ and
the microscopic diffusion coefficient is $D_{12}$. 
The effective gravity is zero in the
${\mathbf \hat{s}}$ direction (from the definition of the
equipotential), and so the third term in parentheses above is zero. 
Also, the thermal diffusion term is typically small compared to
the radiative term
and so the diffusion velocity is determined only by the
radiative forces, and the concentration gradient.

If the mass of
species 1 (the stellar plasma - mostly hydrogen and helium) 
is neglected compared to the mass of species 2 (the diffusing ions),
and ion 
shielding is neglected in the computation of the Debye length, eq.40 of
Paquette \etal\ (1986) becomes
\begin{equation}
D_{12} = \frac{3(2kT)^{5/2} m_p^{1/2}}{16 \rho \pi^{1/2} Z_i^2 e^4
\Lambda_i}
\end{equation}
where
\begin{equation}
\Lambda_i = {\rm ln} \left[ 1 +
\frac{16(kT)^3 m_p}{4\pi Z_i^2 e^6 \rho } \right]
\end{equation}
where $\rho$ is the density.
To obtain a simple expression for the diffusion velocity, an initially
chemically homogeneous
star is assumed, so that $\nabla c = 0$ and eq.9 are good
representations. The diffusion velocity is then 
\begin{equation}
v_s = -2.2\times 10^{-9}
\frac{T_4^\frac{1}{2} T_{e\;4}^4 }
{\rho \! \left( 1 + \frac{c}{c_0} \right)^\frac{1}{2}} \!
\left( \frac{\rstar}{r} \right)^2 \!\!\!
\frac{\Phi_i}{Z_i^2 \Lambda_i} \!
\left( \frac{r}{2} \frac{d{\rm ln}f}{ds} \right)
\end{equation}
This expression will overestimate the drift velocity as soon as a
concentration gradient is established, as then $\nabla c \neq 0$ in
eq.11. However, with these points in mind, eq.14 is used as an
estimate for the velocity in eq.4. 

\section{Numerical estimates of the diffusion velocity}

Here the typical drift velocity is calculated for two model stars.
The stellar models have been provided by M. Salaris (private
communication) and are coomputed according to the evolution code 
described in Salaris \etal\ (1997) and references therein.
The two models correspond to (2.0\msun, 1.8\rsun $T_{\rm eff} = 8700$K) and 
(5.0\msun, 3.0\rsun $T_{\rm eff} = 16500$K) main sequence stars. 
These models are computed with no rotation and solar
composition. Although below the models are used to represent non-solar
composition rotating stars, it is unlikely that this slight
inconsistency will introduce a large error. 

The calculation of the atomic parameters $\Phi_i$ and $\Psi_i$ are
lengthy and cumbersome, and so the parameters given by Alecian \etal\
(1993) for 
iron are assumed (their table 4), derived using Opacity Project data.
These are for the states
Fe{\footnotesize {IX}}-- Fe{\footnotesize {XVII}}. 
For a given temperature in the model stars, the atomic
parameters used are interpolated between those for the dominant, and
next dominant ions (the temperatures at which ions are dominant is
given in eq.3 of Alecian \etal\ 1993). 
These temperatures also give the applicable range used in the
calculation : $1.95\times 10^4K < T < 6.0\times 10^5K$ to cover all
the ions used. 

Although the models are calculated with solar metallicity, the drift
velocity $v_s$ is calculated for two values of iron abundance:
solar abundance $c = 3\times 10^{-5}$ and $10^{-2}$ solar. Although
strictly this is inconsistent, the error associated in doing so will
be very small.

In fig.2, the {\em radial} radiative acceleration is shown for the two
stars with solar iron abundance, and should be compared with fig.4 of
Alecian \etal\ (1993). It can be seen that the interpolation used here
to calculate the atomic parameters is not introducing any large
errors in the calculation.
\begin{figure}
\begin{picture}(100,270)
\put(0,0){\includegraphics{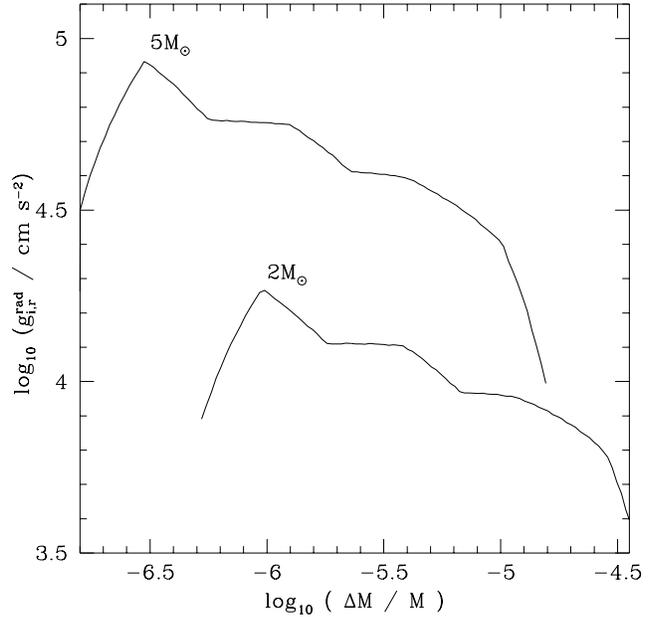}}
\end{picture}
\caption{Numerical estimates of the radial radiative acceleration
$g^{\rm rad}_{i,r}$ for the two model stars.}
\end{figure}

Fig.3 shows the value of $v_s$ as a function of the fractional mass
exterior to that point $\Delta M / M$ for the two models. 
The stars is assumed to rotate at 0.7 of their break up
velocities (0.89 of its break-up angular velocities), equal to 
$v_{\rm rot} = 260\kms$ and 320\kms\ respectively for the 2\msun and
5\msun stars.
The top panel of fig.3 corresponds to the 5\msun star, and 
the lower to the 2\msun star. In each panel, the lines refer to 
$\theta = 10^\circ$, $30^\circ$, $50^\circ$ and
$70^\circ$ with the uppermost line corresponding to $70^\circ$
decreasing downwards.
The solid and dotted lines correspond to iron abundance of solar and
$10^{-2}$solar respectively. 
\begin{figure}
\begin{picture}(100,480)
\put(0,0){\includegraphics{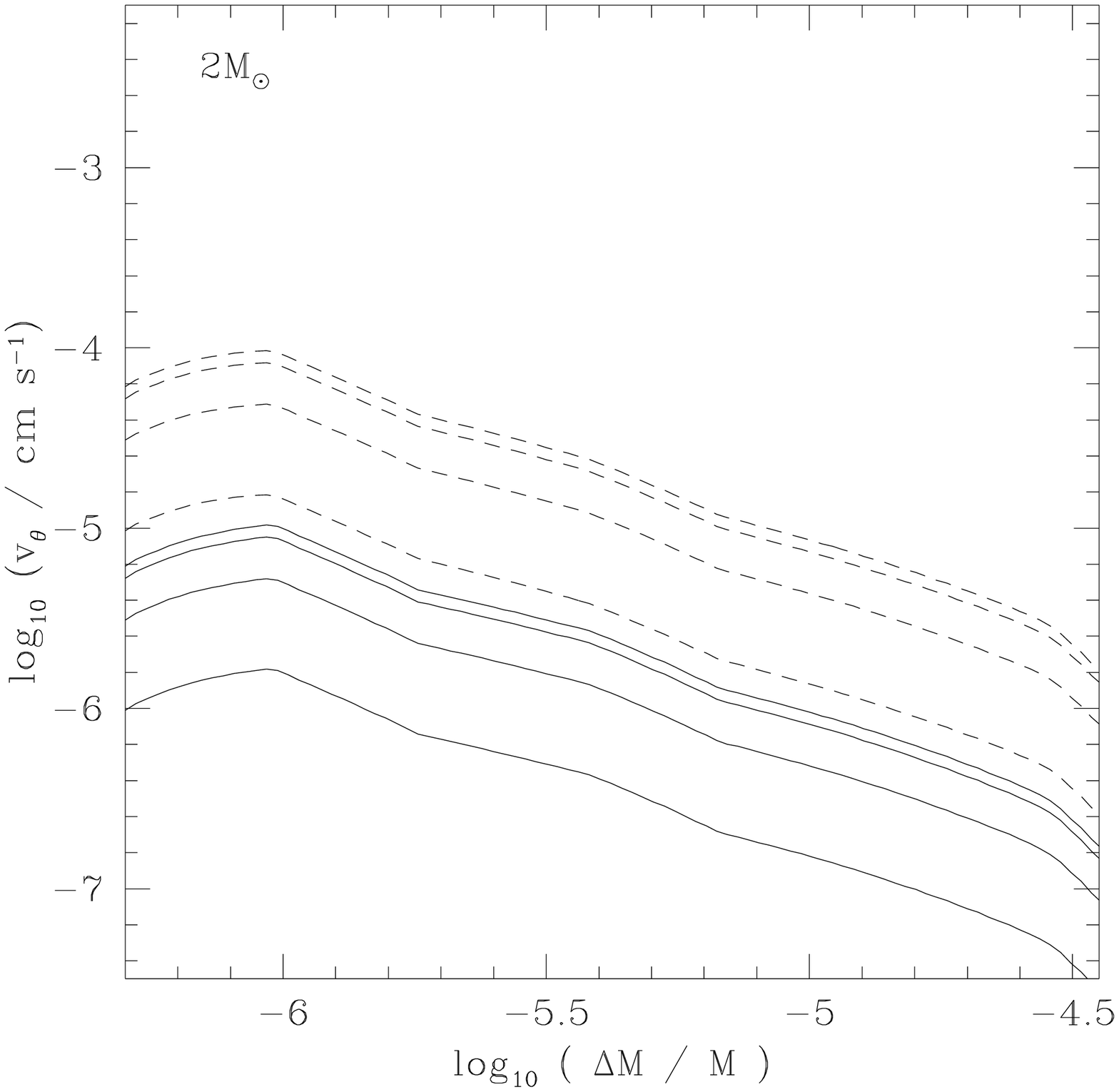}}

\put(0,0){\includegraphics{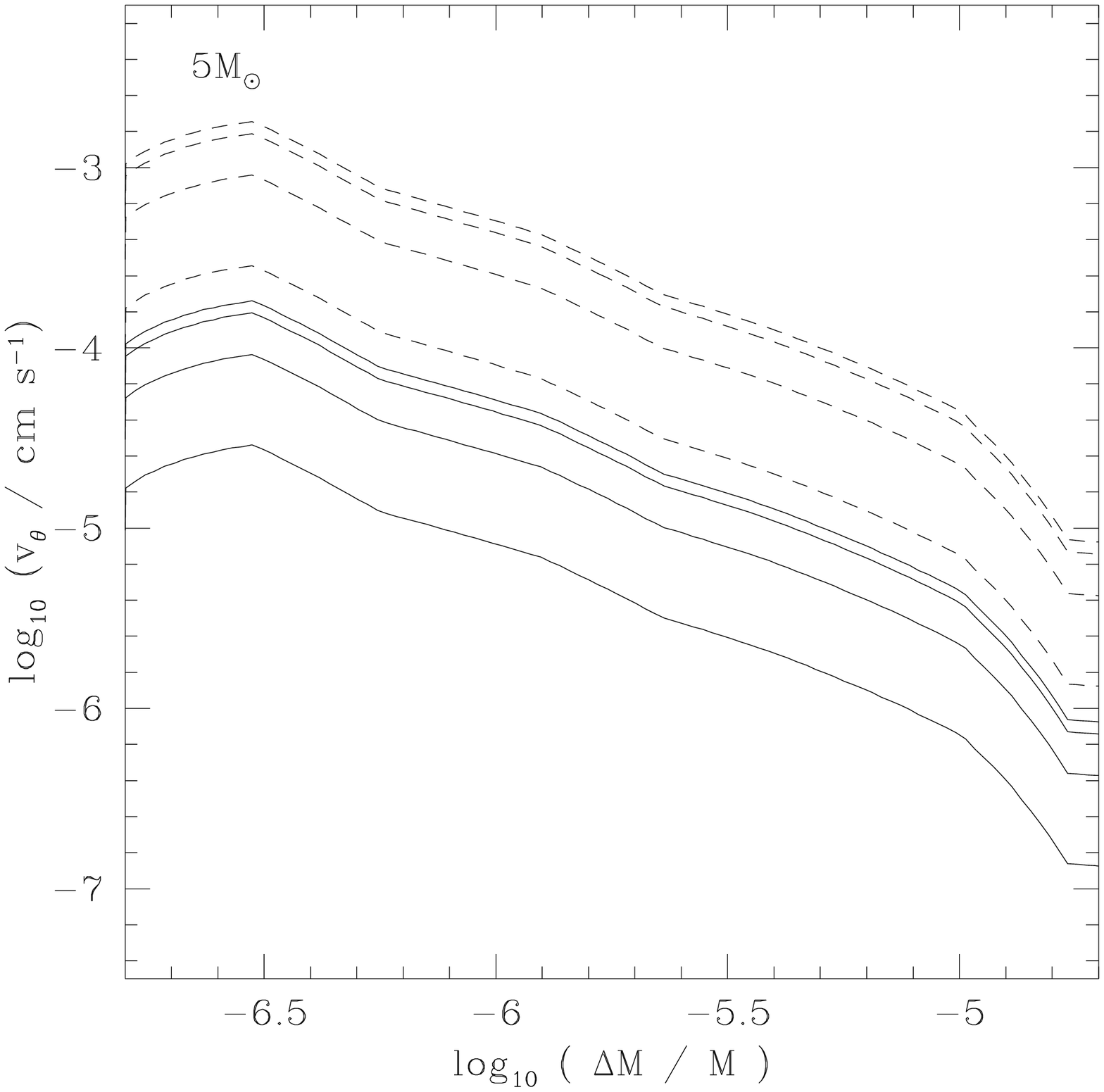}}
\end{picture}
\caption{Numerical estimates of $v_s$ from eq.14. 
The four lines in each panel refer to velocities at 
$\theta = 10^\circ$, $30^\circ$, $50^\circ$ and
$70^\circ$ with the uppermost line corresponding to $70^\circ$
decreasing downwards. The solid (dashed) lines correspond to solar
($10^{-2}$ solar) abundance of iron. }
\end{figure}

In three of the four models, the drift velocity exceeds $10^{-5}$cm
s$^{-1}$ in 
the outer layers at some latitude. This indicates that there may be
regions in the envelope where iron may start to diffuse round the star.
For the 5\msun star the outer $10^{-5.5}$ ($10^{-4.7}$) of the solar
($10^{-2}$ solar) composition model may suffer latitudinal drift of
iron. For the 2\msun star, at $10^{-2}$ solar composition the situation is
similar -- $v_s \gtappeq 10^{-5}$cm s$^{-1}$ in the outer $10^{-5}$ of
the star.
However for solar composition, there may be no effect as the
radiative acceleration on iron is smaller. In this case the turbulence
may homogenise the star faster than the metals may drift due to the radiation.
Taken at face value, fig.3 indicates that stars rotating at
significant fractions of their break-up velocity may be prone to
latitudinal metallic drift in their outer envelopes.


\section{Timescales}

Is is likely that a star may come into
equilibrium? This clearly depends on the timescales over which the
relevant processes act.
Here the important timescales are 
(i) the diffusion time (time taken for the ions to diffuse round that
star), (ii) the main-sequence lifetime of the star, (iii) the lifetime
of each layer (if mass-loss is present).

\subsection{Diffusion, stellar and mass-loss timescales}
For an asymmetry to occur in the elemental distribution along an
equipotential, the metals must diffuse a significant distance from
their initial latitude to 
the equator. 
This therefore defines a drift timescale $\tau_D$, dependent on
latitude $\pi/2 - \theta$ , 
which is that over which a latitudinal asymmetry in the metals will
be generated: 
\begin{equation}
\tau_D(\theta) \sim \frac{\rstar (\pi/2 - \theta) }{v_s}
\end{equation} 

The star only has a finite time in which to allow metals diffuse
around its equipotentials. 
The main sequence lifetimes for these models are 
$\tau_{MS} = 9.5\times 10^8$yr and  $7.9\times 10^7$yr for the 2\msun and
5\msun star respectively.

The mass-loss timescale within the envelope depends on the mass
exterior to that point ($\Delta M$) and the mass loss rate \mdot. If 
the mass-loss rate
\mdot\ is a function of azimuthal position $\theta$ for rotating stars
(for line-driven winds see Friend \& Abbott 1982 and for
radiation-driven dusty winds see Dorfi \& H\"{o}fner 1996),
then the mass-loss timescale also varies with $\theta$. 
However, as a first approach $\tau_M(\theta)$ is taken to be
independent of $\theta$:
\begin{equation}
\tau_M \sim \frac{\Delta M}{\mdot}.
\end{equation}

\subsection{Where can the asymmetry exist?}
In order for any significant metallic asymmetry to be generated, the
diffusion timescale must be less than the mass-loss timescale.
This
ensures that the diffusion process has enough time to act before that
layer of the star is lost in an outflow. The diffusion timescale must also
be shorter than the typical main-sequence timescale else no significant
asymmetry will build up over the lifetime of the star.
Therefore, chemical abundance variations may be generated in regions
for which the two inequalities
\begin{equation}
\tau_D \ltappeq \tau_M, \;\;\;
\tau_D(\theta) \ltappeq \tau_{MS}
\end{equation}
are satisfied. 
\begin{figure}
\begin{picture}(100,490)
\put(0,0){\includegraphics{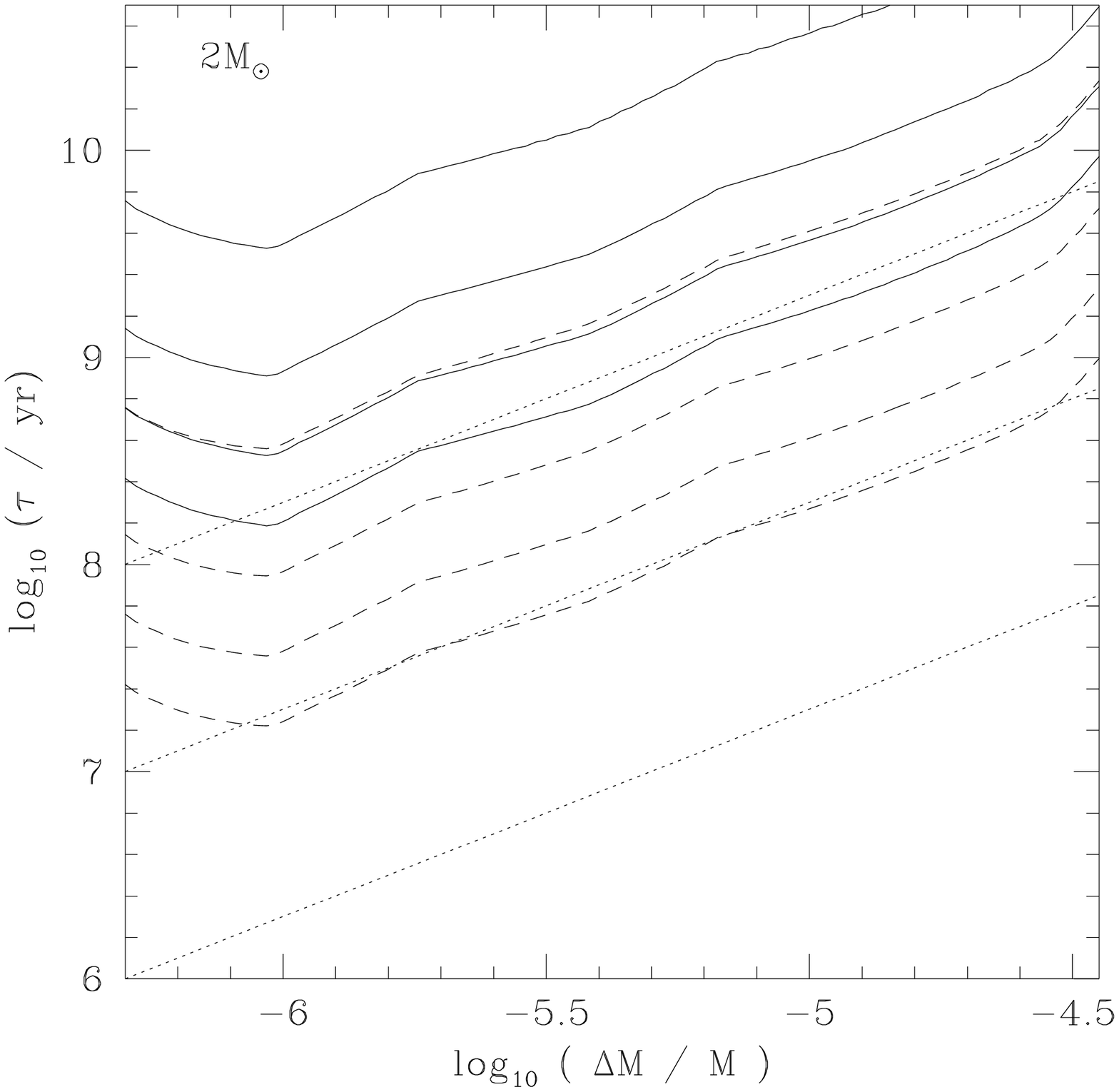}}

\put(0,0){\includegraphics{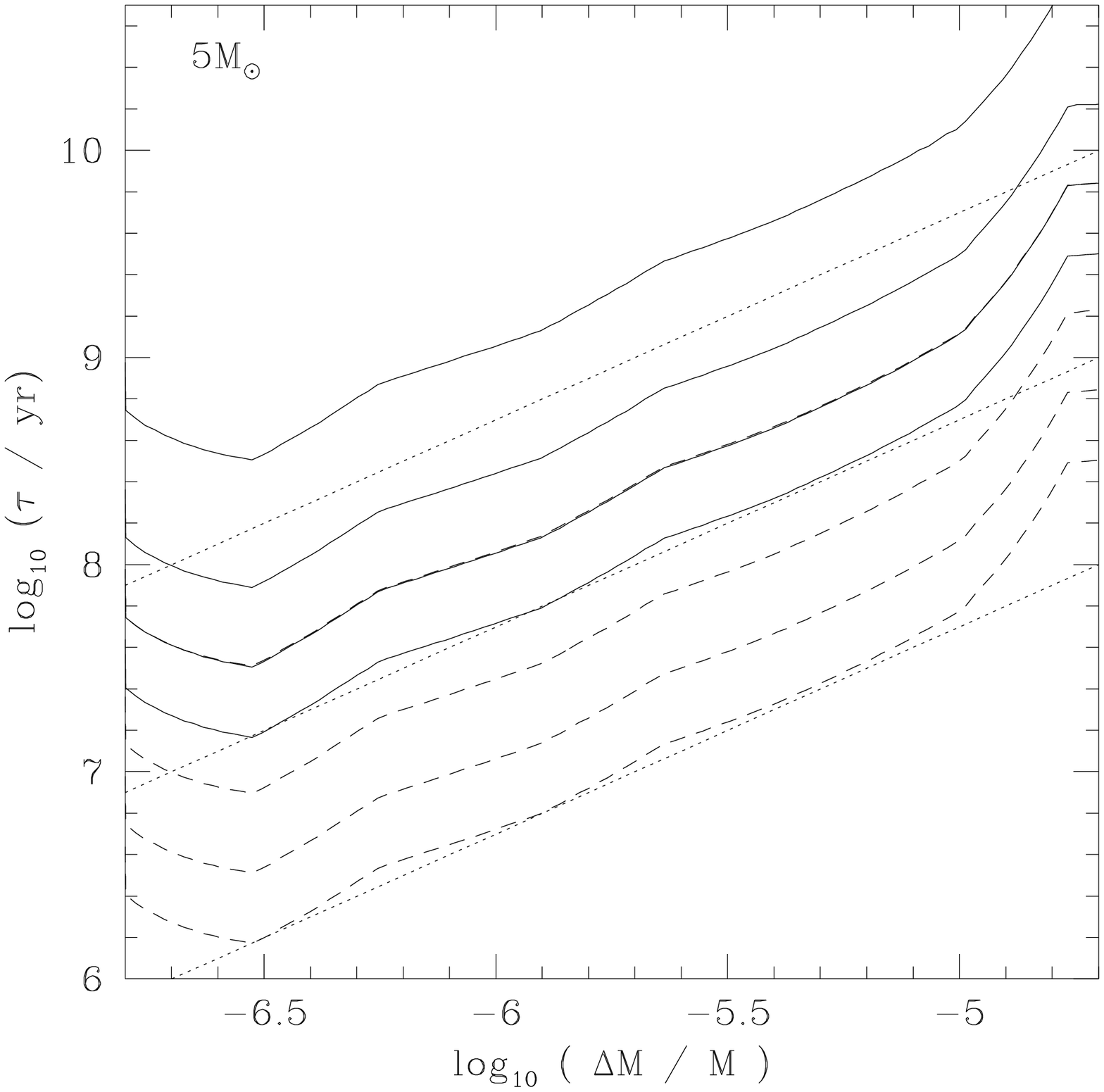}}
\end{picture}
\caption{ The timescales in the star: the diffusion
timescale around the star for $\theta = 10^\circ$, $30^\circ$, $50^\circ$ and
$70^\circ$ with the lower line corresponding to $70^\circ$
decreasing upwards. The solid (dashed) lines correspond to solar
($10^{-2}$ solar) abundance of iron.
The dotted lines are the
mass-loss timescales corresponding to mass-loss rates of $10^{-12}$\msunyr
(lower), $10^{-13}\msunyr, \ 10^{-14}$\msunyr (upper).}
\end{figure}

The timescales are displayed in fig.4. The lines are the diffusion
timescale around the star for $\theta = 10^\circ$, $30^\circ$, $50^\circ$ and
$70^\circ$ with the lower line corresponding to $70^\circ$
decreasing upwards. 
The solid lines are for solar iron abundance, and the dotted lines are
for $10^{-2}$ solar abundance.
The oblique dashed lines are the mass loss timescale,
corresponding to mass-loss rates of $10^{-12}$\msunyr
(lower), $10^{-13}\msunyr, \ 10^{-14}$\msunyr (upper).
Fig.4 should be interpreted as follows: the region to the right and
lower than the mass-loss timescale lines may, if the
diffusion velocity is larger than the diffusive velocity ($\sim$ the
meridional velocity $u_r$ see \S2) develop metallic abundance
gradients around an equipotential. Therefore, the 5\msun model may
have enough time to develop a significant abundance gradient across 
most of the meridian in the outer regions is the mass-loss rate is
$\sim 10^{-13}$\msunyr.

Fig.4 suggests that during the main sequence lives of 
the $10^{-2}$solar abundance 2\msun and both 5\msun stars, 
the drift velocity due to the anisotropic flux in the star $v_s$
is large enough that a latitudinal abundance gradient may be generated. 
However, this statement must be confined to stars which are
rotating at significant fractions of break-up, and only applies to 
bands near to the equator.

Although this effect may produce
metallic drift, it seems from fig.4 unlikely that the whole of an
equipotential $s$ will be involved. It appears that only regions close
to the equator will be effected and consequently the metallic
distribution will {\em not} come into equilibrium during the star's
main-sequence lifetime.
The distribution of metals is therefore difficult to calculate --
indeed the dynamical modelling of the distribution is out of the scope
of this paper and is flagged for further study.
 
\section{Discussion}

The results of the previous sections are startling. It has been found
that in bands near the equator metals may drift toward the equatorial
plane due to the anisotropic radiation field created when the star
rotates. 
However, it is noted that if the mass-loss rate in the equatorial
regions is strongly dependent on the metallicity, then the enhanced
mass-loss rate will reduce the time that a given layer is bound to the star.
Therefore the metallicity enhancement will not be able to build
up to such a large level (as estimated below). 
In this case then a new equilibrium will be
approached in which the magnitude of the metallicity enhancement is
controlled by the enhanced mass-loss. 
This aspect of the problem is currently being investigated -- the
homogenising effect of the outer convection zone in the models has,
so far, been neglected. This may homogenise the metallicity fast
enough that no effect on the mass-loss rate may be observed.

The drift will slow and eventually stop when the logarithmic
concentration gradient becomes comparable with the radiative force
(see eq.11):
\begin{equation}
\frac{d{\rm ln}c}{d\theta} \approx A_i \frac{m_p g^{\rm rad}_{i, s}}{kT} r
\end{equation}
(again note this is an approximation as $\hat{\mathbf s}$ and
$\hat{\mathbf \theta}$ are not parallel).
As an example, the temperature is set to $2.5\times 10^5$K
corresponding to $\Delta M /M \sim 10^{-6}$ for both stars. 
The radii of the stars of $r = 3.0\rsun$ (5\msun)
and 1.8\rsun (2\msun).
From fig.2 the radial
radiative force is $\approx 10^{4.7}$ ($10^{4.1}$) for the 5\msun
(2\msun) star. If it is assumed that the radiative force around the
star $g^{\rm rad}_{i, s}$ is a tenth of this (see fig.1)
then the right-hand side of eq.18 is 1360 and 205 for the 5\msun and
2\msun stars respectively. 
These are clearly large numbers -- indicative of a very large abundance
build-up in the equatorial plane. However as shown in \S6 it is 
unlikely that the whole of the meridian will come into equilibrium. 
It may be surmised, though, that significant abundance inhomogeneities
can be generated (without being thwarted by the concentration
gradient) through rotation. 

In the course of this paper, radial motions due to radiative
acceleration or gravitational settling have been neglected.
It is likely that a radial component of drift velocity is
present as well as the meridional component focussed on
here. Therefore, the motion of heavy element ions will not be solely be
meridional -- if the radial radiative acceleration exceedes gravity,
then the drift will be towrd the equator {\em and} toward the surface.

It is difficult to accurately assess the effects of large latitudinal
abundance gradients on the structure of a
star. As soon as the abundances change at a given point, then the
flux distribution will also change along with the local convective
stability criterion.
Convection smooths out the abundance overdensity, and provides some
feedback to the ionic build up. 
It is posssible that this feedback will regulate the ionic drift,
modifying the structure of parts of the star.

Let us now consider the evolution of the outer envelope of the star during
the main sequence. As the very outer layers are lost in a wind, then a given
layer becomes slightly more diffuse and the layer moves to lower
$\Delta M / M$ (which will promote latitudinal
ion drift). However, as the layer becomes 
convectively unstable then it is quickly made homogeneous.
The presence of extra radiation-blocking ions in fact make
the layer more unstable to convection, and so the equatorial regions
of rotating stars can be expected to have a more extensive convective region. 
Due to this rapid convective homogenisation,
it is very unlikely then that this latitudinal drift will be observed in
the photosphere of the star during its main sequence lifetime 
(the possibility of a star having different chemical compositions on
the same equipotential has already been considered using radial
diffusion and magnetic fields by Michaud, M\'{e}gessier \& Charland
1981). The possibility of observing a direct manifestation of
latitidinal abundance variation in post main-sequence phases of
evolution is currently under study. Here the mass-loss rate may be
large enough to prevent complete convective chemical homogenisation of
regions which have generated abundance gradients during the main sequence.

It is clear that as the timescales over which a significant abundance
gradient may be generated in parts of the star are similar to either mass-loss
timescales or main-sequence lifetimes. Therefore the evolution of the
star needs to be taken into account whilst considering the abundance
time dependence -- something which is not attempted here.
Although difficult due to necessarily 2D nature of the calculation, 
it appears that time dependence of the abundance
distribution this needs to be included in models for rotating stars.

\section{Conclusion}

It has been shown that rotation may indirectly have a large effect on
the latitudinal abundance distribution of metals via gravity darkening.
Indeed, the effect is possibly active in all the model stars considered
here. Stars of low metallicity are particularly prone to this effect.
If the mass loss of the star is strongly metallicity dependent, then
the effect in the outer layers may be curtailed somewhat. This may
lead to an extra mass-loss rate enhancement in the equatorial plane.

Although several major points regarding the effects of meridional
diffusion of ions toward the equator have simply been touched on, 
the principle of metallic latitudinal distribution
asymmetries has at least been put on quantitative basis.

\section*{Acknowledgements}
I thank Drs. M. Salaris, P. Podsiadlowski and J. Drew for
discussions regarding this work, and also the referee, A. Maeder, for
constructive comments on the paper.
JMP is supported by a PPARC postdoctoral research assistantship. 


{}


\end{document}